# Plasmonic mode converter for controlling optical impedance and nanoscale light-matter interaction

**Yun-Ting Hung,[1] Chen-Bin Huang,[2] and Jer-Shing Huang[1,3*]**

[1]*Department of Chemistry, National Tsing Hua University, Hsinchu 30013, Taiwan*
[2]*Institute of Photonics Technologies, National Tsing Hua University, Hsinchu 30013, Taiwan*
[3]*Frontier Research Center on Fundamental and Applied Sciences of Matters, National Tsing Hua University, Hsinchu 30013, Taiwan*
[*]*jshuang@mx.nthu.edu.tw*

**Abstract:** To enable multiple functions of plasmonic nanocircuits, it is of key importance to control the propagation properties and the modal distribution of the guided optical modes such that their impedance matches to that of nearby quantum systems and desired light-matter interaction can be achieved. Here, we present efficient mode converters for manipulating guided modes on a plasmonic two-wire transmission line. The mode conversion is achieved through varying the path length, wire cross section and the surrounding index of refraction. Instead of pure optical interference, strong near-field coupling of surface plasmons results in great momentum splitting and modal profile variation. We theoretically demonstrate control over nanoantenna radiation and discuss the possibility to enhance nanoscale light-matter interaction. The proposed converter may find applications in surface plasmon amplification, index sensing and enhanced nanoscale spectroscopy.

**References and links**

1. E. Ozbay, "Plasmonics: Merging photonics and electronics at nanoscale dimensions," Science **311**, 189-193 (2006).
2. J. A. Schuller, E. S. Barnard, W. Cai, Y. C. Jun, J. S. White, and M. L. Brongersma, "Plasmonics for extreme light concentration and manipulation," Nat. Mater. **9**, 193-204 (2010).
3. D. E. Chang, A. S. Sørensen, P. R. Hemmer, and M. D. Lukin, "Quantum optics with surface plasmons," Phys. Rev. Lett. **97**, 053002 (2006).
4. D. E. Chang, A. S. Sorensen, E. A. Demler, and M. D. Lukin, "A single-photon transistor using nanoscale surface plasmons," Nat. Phys. **3**, 807-812 (2007).
5. L. Novotny and N. van Hulst, "Antennas for light," Nat. Photonics **5**, 83-90 (2011).
6. P. Biagioni, J.-S. Huang, and B. Hecht, "Nanoantennas for visible and infrared radiation," Rep. Prog. Phys. **75**, 024402 (2012).
7. J.-S. Huang, T. Feichtner, P. Biagioni, and B. Hecht, "Impedance matching and emission properties of nanoantennas in an optical nanocircuit," Nano Lett. **9**, 1897-1902 (2009).
8. J. Wen, S. Romanov, and U. Peschel, "Excitation of plasmonic gap waveguides by nanoantennas," Opt. Express **17**, 5925-5932 (2009).
9. J.-S. Huang, D. V. Voronine, P. Tuchscherer, T. Brixner, and B. Hecht, "Deterministic spatiotemporal control of optical fields in nanoantennas and plasmonic circuits," Phys. Rev. B **79**, 195441 (2009).
10. P. M. Krenz, R. L. Olmon, B. A. Lail, M. B. Raschke, and G. D. Boreman, "Near-field measurement of infrared coplanar strip transmission line attenuation and propagation constants," Opt. Express **18**, 21678-21686 (2010).
11. M. Schnell, P. Alonso Gonzalez, L. Arzubiaga, F. Casanova, L. E. Hueso, A. Chuvilin, and R. Hillenbrand, "Nanofocusing of mid-infrared energy with tapered transmission lines," Nat. Photonics **5**, 283-287 (2011).
12. J. Wen, P. Banzer, A. Kriesch, D. Ploss, B. Schmauss, and U. Peschel, "Experimental cross-polarization detection of coupling far-field light to highly confined plasmonic gap modes via nanoantennas," Appl. Phys. Lett. **98**, 101109-101103 (2011).

13. N. Yang, Y. Tang, and A. E. Cohen, "Spectroscopy in sculpted fields," Nano Today **4**, 269-279 (2009).
14. S. Berweger, J. M. Atkin, R. L. Olmon, and M. B. Raschke, "Light on the tip of a needle: Plasmonic nanofocusing for spectroscopy on the nanoscale," J. Phys. Chem. Lett. **3**, 945-952 (2012).
15. J. C. Weeber, J. R. Krenn, A. Dereux, B. Lamprecht, Y. Lacroute, and J. P. Goudonnet, "Near-field observation of surface plasmon polariton propagation on thin metal stripes," Phys. Rev. B **64**, 045411 (2001).
16. S. I. Bozhevolnyi, V. S. Volkov, E. Devaux, and T. W. Ebbesen, "Channel plasmon-polariton guiding by subwavelength metal grooves," Phys. Rev. Lett. **95**, 046802 (2005).
17. S. I. Bozhevolnyi, V. S. Volkov, E. Devaux, J.-Y. Laluet, and T. W. Ebbesen, "Channel plasmon subwavelength waveguide components including interferometers and ring resonators," Nature **440**, 508-511 (2006).
18. E. Verhagen, M. Spasenović, A. Polman, and L. Kuipers, "Nanowire plasmon excitation by adiabatic mode transformation," Phys. Rev. Lett. **102**, 203904 (2009).
19. S. Zhang, H. Wei, K. Bao, U. Håkanson, N. J. Halas, P. Nordlander, and H. Xu, "Chiral surface plasmon polaritons on metallic nanowires," Phys. Rev. Lett. **107**, 096801 (2011).
20. C. Rewitz, T. Keitzl, P. Tuchscherer, J.-S. Huang, P. Geisler, G. Razinskas, B. Hecht, and T. Brixner, "Ultrafast plasmon propagation in nanowires characterized by far-field spectral interferometry," Nano Lett. **12**, 45-49 (2012).
21. S. A. Maier, *Plasmonics: Fundamentals and applications* (Springer, Berlin, 2007).
22. J. A. Dionne, H. J. Lezec, and H. A. Atwater, "Highly confined photon transport in subwavelength metallic slot waveguides," Nano Lett. **6**, 1928-1932 (2006).
23. J. Kern, S. Grossmann, N. V. Tarakina, T. Häckel, M. Emmerling, M. Kamp, J.-S. Huang, P. Biagioni, J. C. Prangsma, and B. Hecht, "Atomic-scale confinement of optical fields," arXiv, http://arxiv.org/abs/1112.5008v2, (2011).
24. D. J. Bergman and M. I. Stockman, "Surface plasmon amplification by stimulated emission of radiation: Quantum generation of coherent surface plasmons in nanosystems," Phys. Rev. Lett. **90**, 027402 (2003).
25. M. A. Noginov, V. A. Podolskiy, G. Zhu, M. Mayy, M. Bahoura, J. A. Adegoke, B. A. Ritzo, and K. Reynolds, "Compensation of loss in propagating surface plasmon polariton by gain in adjacent dielectric medium," Opt. Express **16**, 1385-1392 (2008).
26. I. De Leon and P. Berini, "Amplification of long-range surface plasmons by a dipolar gain medium," Nat. Photonics **4**, 382-387 (2010).
27. M. C. Gather, K. Meerholz, N. Danz, and K. Leosson, "Net optical gain in a plasmonic waveguide embedded in a fluorescent polymer," Nat. Photonics **4**, 457-461 (2010).
28. A. V. Krasavin, T. P. Vo, W. Dickson, P. d. M. Bolger, and A. V. Zayats, "All-plasmonic modulation via stimulated emission of copropagating surface plasmon polaritons on a substrate with gain," Nano Lett. **11**, 2231-2235 (2011).
29. P. Berini and I. De Leon, "Surface plasmon-polariton amplifiers and lasers," Nat. Photonics **6**, 16-24 (2012).
30. S. Kühn, U. Håkanson, L. Rogobete, and V. Sandoghdar, "Enhancement of single-molecule fluorescence using a gold nanoparticle as an optical nanoantenna," Phys. Rev. Lett. **97**, 017402 (2006).
31. FDTD solutions, Lumerical Solutions Inc., Vancouver, Canada. http://www.lumerical.com/.
32. P. B. Johnson and R. W. Christy, "Optical constants of the noble metals," Phys. Rev. B **6**, 4370-4379 (1972).
33. L. Novotny and C. Hafner, "Light propagation in a cylindrical waveguide with a complex, metallic, dielectric function," Phys. Rev. E **50**, 4094-4106 (1994).
34. J. Takahara, S. Yamagishi, H. Taki, A. Morimoto, and T. Kobayashi, "Guiding of a one-dimensional optical beam with nanometer diameter," Opt. Lett. **22**, 475-477 (1997).
35. M. I. Stockman, "Nanofocusing of optical energy in tapered plasmonic waveguides," Phys. Rev. Lett. **93**, 137404 (2004).
36. K. F. MacDonald, Z. L. Samson, M. I. Stockman, and N. I. Zheludev, "Ultrafast active plasmonics," Nat. Photonics **3**, 55-58 (2009).
37. Y. Liu, T. Zentgraf, G. Bartal, and X. Zhang, "Transformational plasmon optics," Nano Lett. **10**, 1991-1997 (2010).
38. E. Verhagen, L. Kuipers, and A. Polman, "Plasmonic nanofocusing in a dielectric wedge," Nano Lett. **10**, 3665-3669 (2010).
39. Y. Gao, Q. Gan, Z. Xin, X. Cheng, and F. J. Bartoli, "Plasmonic Mach–Zehnder interferometer for ultrasensitive on-chip biosensing," ACS Nano **5**, 9836-9844 (2011).
40. P. Nordlander, C. Oubre, E. Prodan, K. Li, and M. I. Stockman, "Plasmon hybridization in nanoparticle dimers," Nano Lett. **4**, 899-903 (2004).
41. J.-S. Huang, J. Kern, P. Geisler, P. Weinmann, M. Kamp, A. Forchel, P. Biagioni, and B. Hecht, "Mode imaging and selection in strongly coupled nanoantennas," Nano Lett. **10**, 2105-2110 (2010).
42. P. J. Schuck, D. P. Fromm, A. Sundaramurthy, G. S. Kino, and W. E. Moerner, "Improving the mismatch between light and nanoscale objects with gold bowtie nanoantennas," Phys. Rev. Lett. **94**, 017402 (2005).
43. S. Chandrasekhar, A. S. Vengurlekar, V. T. Karulkar, and S. K. Roy, "Temperature, light intensity and microstructure dependence of the refractive index of polycrystalline silicon films," Thin Solid Films **169**, 205-212 (1989).
44. E. M. True and L. McCaughan, "Large nonresonant light-induced refractive-index changes in thin films of amorphous arsenic sulfide," Opt. Lett. **16**, 458-460 (1991).

45. N. Large, M. Abb, J. Aizpurua, and O. L. Muskens, "Photoconductively loaded plasmonic nanoantenna as building block for ultracompact optical switches," Nano Lett. **10**, 1741-1746 (2010).
46. A. Cavalleri, C. Tóth, C. W. Siders, J. A. Squier, F. Ráksi, P. Forget, and J. C. Kieffer, "Femtosecond structural dynamics in $VO_2$ during an ultrafast solid-solid phase transition," Phys. Rev. Lett. **87**, 237401 (2001).
47. M. Seo, J. Kyoung, H. Park, S. Koo, H.-S. Kim, H. Bernien, B. J. Kim, J. H. Choe, Y. H. Ahn, H.-T. Kim, N. Park, Q. H. Park, K. Ahn, and D.-S. Kim, "Active terahertz nanoantennas based on $VO_2$ phase transition," Nano Lett. **10**, 2064-2068 (2010).
48. S. Y. Park and D. Stroud, "Splitting of surface plasmon frequencies of metal particles in a nematic liquid crystal," Appl. Phys. Lett. **85**, 2920-2922 (2004).
49. W. Dickson, G. A. Wurtz, P. R. Evans, R. J. Pollard, and A. V. Zayats, "Electronically controlled surface plasmon dispersion and optical transmission through metallic hole arrays using liquid crystal," Nano Lett. **8**, 281-286 (2008).
50. J. Berthelot, A. Bouhelier, C. Huang, J. Margueritat, G. Colas-des-Francs, E. Finot, J.-C. Weeber, A. Dereux, S. Kostcheev, H. I. E. Ahrach, A.-L. Baudrion, J. Plain, R. Bachelot, P. Royer, and G. P. Wiederrecht, "Tuning of an optical dimer nanoantenna by electrically controlling its load impedance," Nano Lett. **9**, 3914-3921 (2009).
51. W. L. Barnes, A. Dereux, and T. W. Ebbesen, "Surface plasmon subwavelength optics," Nature **424**, 824-830 (2003).
52. A. Guerrero-Martínez, M. Grzelczak, and L. M. Liz-Marzán, "Molecular thinking for nanoplasmonic design," ACS Nano **6**, 3655-3662 (2012).
53. J.-J. Greffet, M. Laroche, and F. Marquier, "Impedance of a nanoantenna and a single quantum emitter," Phys. Rev. Lett. **105**, 117701 (2010).
54. A. M. Boiron, B. Lounis, and M. Orrit, "Single molecules of dibenzanthanthrene in n-hexadecane," J. Chem. Phys. **105**, 3969-3974 (1996).
55. W. E. M. T. Basché, M. Orrit, U. P. Wild, *Single-molecule optical detection, imaging and spectroscopy* (Wiley-VCH, Munich, 1997).
56. J.-S. Huang, V. Callegari, P. Geisler, C. Brüning, J. Kern, J. C. Prangsma, X. Wu, T. Feichtner, J. Ziegler, P. Weinmann, M. Kamp, A. Forchel, P. Biagioni, U. Sennhauser, and B. Hecht, "Atomically flat single-crystalline gold nanostructures for plasmonic nanocircuitry," Nat. Commun. **1**, 150 (2010).
57. D. W. Pohl, S. G. Rodrigo, and L. Novotny, "Stacked optical antennas," Appl. Phys. Lett. **98**, 023111-023113 (2011).

## 1. Introduction

Surface plasmon polaritons (SPPs) are promising for the realization of optical nanocircuits [1] and provide possible solutions to control light-matter interaction on the nanometer scale [2], paving the way to quantum optics with surface plasmons [3, 4]. Recently, prototype nanocircuits using gap nanoantennas [5, 6] in connection with plasmonic two-wire transmission lines (TWTLs) have been theoretically proposed [7-9] and experimentally studied [10-12], showing a realizable approach to the manipulation of electromagnetic field for enhanced spectroscopy [13, 14].

In order to guide the localized optical fields, various plasmonic nanowaveguides have been proposed and studied [15-20]. Two commonly used waveguides are metal nanowires [15, 18-20] and nanogrooves in metallic films [16, 17], representing the insulator-metal-insulator (IMI) and metal-insulator-metal (MIM) waveguides [21], respectively. Plasmonic TWTLs consisting of two parallel metallic nanowires with a nanosized dielectric gap may function both as IMI and MIM waveguides depending on the phase difference $\Delta\Phi$ between the displacement currents on the wires. With in-phase displacement currents ($\Delta\Phi = 0$), the charge distribution is symmetric across the gap, hence the electric field hardly enters the gap and the power is mostly guided through the outer surface of the two wires. Such a spatially less confined mode has a transverse magnetic (TM) character and is analogous to the guided fundamental mode on an IMI waveguide. With out-of-phase currents ($\Delta\Phi = \pm\pi$), the opposite charges result in a highly confined field in the dielectric nanogap [7, 22, 23] and the polarization of the electric field is well transverse to the propagation direction, *i.e.* transverse electric (TE).

In view of power and signal transmission in a functional nanocircuit, the TM mode on a TWTL is more competent since it has higher group velocity and longer propagation length thanks to the less-confined field distribution. For the compensation of inherent loss in plasmonic nanocircuits, such large and loose modal distribution also allows for collecting energy from more excited gain materials [24-29]. On the other hand, for enhancing nanoscale light-matter interaction, guided TE mode on a TWTL is more promising since it provides extreme spatial confinement, large intensity enhancement and well-defined polarization. Considering the interaction between a polarized electric field $E$ and a dye molecule, the excitation efficiency is proportional to $|\langle e|E \cdot D|g\rangle|^2$, with $D$ being the dipole moment operator and $|e\rangle$ and $|g\rangle$ being the excited and ground state of the dye molecule, respectively [30]. It is obvious that both the intensity and the polarization of the electric field play important roles in the light-matter interaction. To realize functional plasmonic nanocircuits, single quantum systems inside the nanogap are of great interest since they may serve as nanosized transistors for surface plasmons [4]. In order to fill the requirements of various circuit functions and to manipulate the light-matter interaction, it is, therefore, important to have the ability to control the guided modes such that the impedance matches to that of nearby nanoobjects. Figure 1 shows the schematic of the proposed mode converter and illustrates the idea of manipulating the nanoscale light-matter interaction by mode conversion. The guided TM mode with in-phase currents is injected from the left side and is converted by the mode converter into a highly concentrated TE mode in the dielectric nanogap. While the guided TM mode exhibits very low field intensity in the gap and bypasses the quantum systems, the guided TE mode is well-polarized and highly concentrated inside the nanogap and therefore strongly interacts with quantum systems integrated in the gap.

Here, we present efficient mode converters for controlling the optical impedance and light-matter interaction in an integrated plasmonic nanocircuit. To achieve deterministic mode conversion, additional phase difference between the guided SPPs on individual wires is introduced by three means, namely by differentiating the path length, the waveguide cross section and the surrounding refractive index. Compared to conventional Mach-Zehnder interferometers, the proposed mode converters rely on strong near-field coupling of guided SPPs instead of interference of guided photonic modes. With the proposed mode converter, we demonstrate improved impedance matching and controlled emission from a nanoantenna in a complex nanocircuit. We also discuss possible applications in index sensing and enhancing nanoscale light-matter interaction, and provide realizable fabrication strategies. Our plasmonic mode converter is of great interest for the realization of practical multi-functional nanocircuits and enhanced spectroscopy using nanoscopic enhanced optical field.

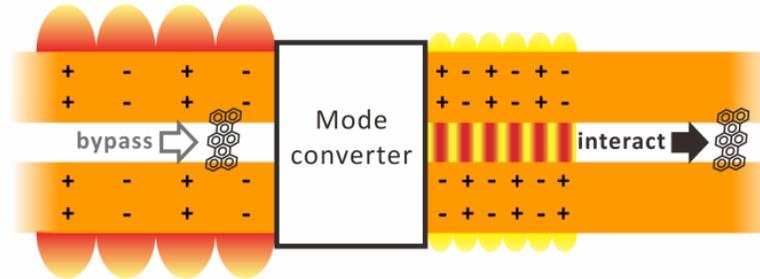

Fig. 1. Schematic of the plasmonic waveguide mode converter for manipulating the nanoscale light-matter interaction. Single terrylene molecules are used as an example for quantum systems integrated in the nanogap with well-aligned permanent dipole moment.

## 2. Analysis on plasmonic waveguiding

Theoretical and numerical analyses on plasmonic single-wire waveguide and TWTL are provided in this section. Our numerical studies are performed using three-dimensional finite-difference time-domain (FDTD) method and eigen-mode solver based on finite-difference frequency-domain (FDFD) method [31] at a fixed frequency of 361.196 THz, corresponding to a wavelength of 830 nm in vacuum. We have chosen the operational wavelength in order to conform to our current equipped laser system for further experiments. However, we emphasize here that the design principle demonstrated in the work is general and not limited to 830 nm. The dielectric function of gold is fitted to experimental values [32] and the minimum mesh size is set to 1 nm$^3$. Nanowires with rectangular cross section are surrounded either by vacuum or by a homogeneous dielectric material. In three dimensional FDTD simulations, all boundaries are set to contain 12 perfectly matched layers and are positioned at least half wavelength away from the structure to avoid absorption of the near field.

*2.1. Waveguiding on a single nanowire*

For a guided mode on a single metallic nanowire, the electric field phasor in the dielectric medium at a position $r$ along the wire has a form of $E_0 e^{i\tilde{k}_\parallel r} e^{i\theta}$, where $E_0$ is the initial amplitude, $\tilde{k}_\parallel = k_\parallel + ik'_\parallel$ is the complex wavevector, $\theta$ denotes the phase shift to a reference time-harmonic field, typically the excitation source, and $i$ is the imaginary unit. The complex wavevector can be expressed as a product of the complex effective index $\tilde{n}_{\text{eff}}$ and the corresponding wave number in free space $k_0$ by

$$\tilde{k}_\parallel = \tilde{n}_{\text{eff}} k_0. \qquad (1)$$

The real part of the wavevector $k_\parallel$ represents the propagation constant, which describes the phase evolution in space and links to the free space wavelength $\lambda_0$ by

$$k_\parallel = \text{Re}\{\tilde{n}_{\text{eff}}\}\frac{2\pi}{\lambda_0}. \qquad (2)$$

The imaginary part $k'_\parallel$ is the damping constant concerning the inherent loss of the plasmonic waveguide and determines the amplitude propagation length $L$ by

$$L = \left[k'_\parallel\right]^{-1} = \left[\text{Im}\{\tilde{n}_{\text{eff}}\}\frac{2\pi}{\lambda_0}\right]^{-1}. \qquad (3)$$

Note that, in real experiment, the measurable quantity is the near-field intensity whose propagation length is further reduced by a factor of 2.

Changing the cross sectional geometry [33-35], permittivity of metal $\varepsilon_m$ [36] or dielectric function of the surrounding medium $\varepsilon_d$ [37-39] can lead to a strong modification of the wavevector of the guided SPPs. Such unique properties of plasmonic waveguides find no counter parts in low-frequency electronic circuitry where metals behave as perfect conductors.

Figure 2(a) shows the propagation constant $k_\parallel$ and the propagation length $\left[k'_\parallel\right]^{-1}$ as a function of the width of wire cross section. The propagation constant decreases and the propagation length increases as the width increases. For thicker wires (width > 60 nm), the propagation constant gradually saturates due to the finite penetration depth, which is about 30 nm for gold at 361.196 THz. A consequence is that the contribution from higher order modes becomes significant for thicker wires [19, 20]. To avoid the complexity, we use only very thin nanowires with rectangular cross section of 30×30 nm$^2$ and consider only the contribution from the 0th-order TM mode on a single gold nanowire [34]. With such a small cross section, the contribution from higher order modes is negligible due to either very short propagation length or very low initial amplitude [7, 19].

Figure 2(b) shows the dependence of the propagation constant and propagation length on the refractive index $n$ of the surrounding medium. As can be seen, both quantities are sensitive to the variation of the surrounding refractive index. The propagation constant exhibits roughly a linear dependence on the refractive index with a slope $s = 23.76$ rad·μm$^{-1}$·RIU$^{-1}$ around 830 nm (RIU, refractive index unit). The slope increases with decreasing the wire cross section and represents the sensitivity to the index change which may find applications in plasmonic sensing [39]. We will later exploit such dependence of wavevector on the refractive index to control the emission of a nanoantenna.

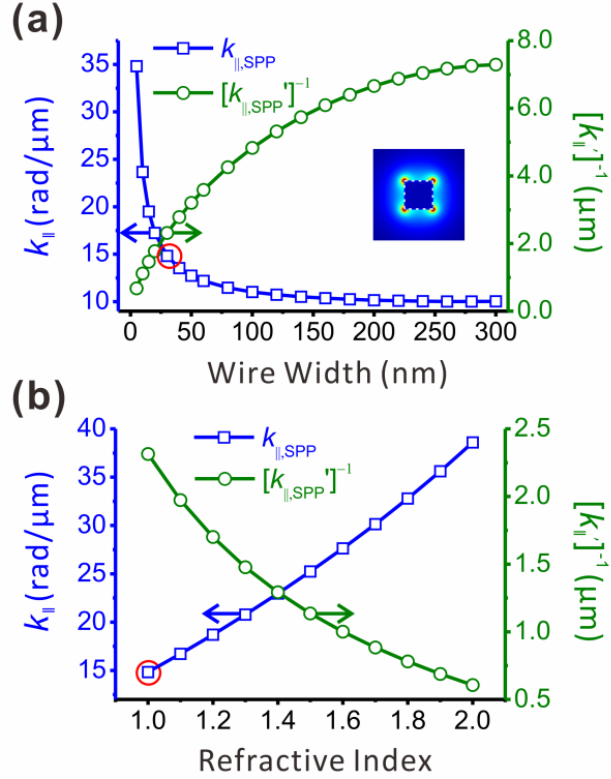

Fig. 2. Propagation constant ($k_\parallel$, blue squares) and the propagation length ($[k_\parallel^{'}]^{-1}$, green circles) of the guided SPPs, i.e. the lowest order TM mode, on a solitary single metallic nanowire as functions of (a) the width of the nanowire and (b) the refractive index $n$ of the surrounding medium. The height of the wire is fixed at 30 nm and the surrounding refractive index is set to 1.0 for (a). The wire cross section is fixed at 30×30 nm$^2$ for (b). The red circles in both plots mark the propagation constant ($k_\parallel = 14.8$ rad/μm) of a nanowire with cross section of 30×30 nm$^2$ embedded in a medium with $n$ = 1.0. The inset shows a representative field intensity profile of the guided 0$^{th}$-order TM mode on a single nanowire.

### 2.2. Waveguiding on a plasmonic TWTL

A nanosized TWTL is essentially a pair of nanowires with their guided SPPs transversely coupled through near-field interaction, similar to mode hybridization between strongly coupled plasmonic nanoparticle dimers [23, 40, 41]. Such coupling results in two distinct guided modes with anti-symmetric and symmetric charge distribution, corresponding to the fundamental TE and TM modes of a plasmonic TWTL, respectively. Figure 3(a) illustrates the wavevector splitting due to the coupling between guided SPPs, as well as the corresponding modal profiles on a TWTL. The wavevector difference between guided TE and TM mode increases as the gap size decreases due to enhanced coupling strength. For guided TE mode, the opposite charges on the two wires result in a highly confined and enhanced field in the dielectric nanogap. The extremely concentrated field in the gap exhibits a well-defined polarization transverse to the propagation direction, which is of great potential for the enhancement of light-matter interaction at the nanoscale [6, 14, 42]. The guided TM mode, on the other hand, shows very low field intensity in the gap and, consequently, does not interact significantly with nanomatter in the gap.

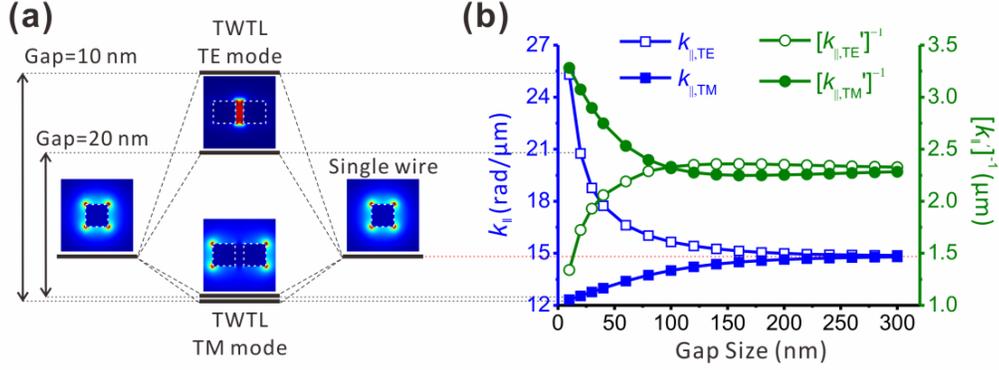

Fig. 3. Coupling of the guided SPPs on individual wires of a nanosized plasmonic TWTL consisting of two gold nanowires (cross section: 30×30 nm$^2$) separated by a nanogap varying from 10 nm to 300 nm. (a) Intensity modal profiles of the guided TE and TM modes on a TWTL as well as that of a solitary single nanowire. The propagation constant at a fixed frequency of 361.196 THz reveals the mode coupling with a gap-dependent splitting of $k_\parallel$, in analogy to the energy splitting observed in strongly coupled nanoantennas [41]. (b) The propagation constant ($k_\parallel$, blue squares) and the propagation length ($\left[k_\parallel'\right]^{-1}$, green circles) of the guided TE (hollow) and TM modes (solid) on a TWTL as a function of the gap size. All modal profiles are normalized to the same intensity color scale with red and blue colors meaning high and low field intensity, respectively.

While the coupling between nanoparticle dimmers are usually observed by the shift in the resonant frequency, the coupling between guided SPPs at a fixed operation frequency is revealed in the wavevector splitting. Such strong coupling is missing for guided photonic modes on dielectric or photonic crystal waveguides and cannot be exploited in conventional optical interferometry, where the signal modulation is a result of pure interference effect. From an experimental point of view, operating the circuit at a constant frequency can be beneficial since variation due to the dispersion of material dielectric function or frequency-dependent polarizability of the quantum systems can be avoided. Figure 3(b) plots the propagation constant and propagation length as a function of the gap size. The splitting of the propagation constant decreases with increasing gap size, indicating a weaker coupling strength for larger gap size. For gap size larger than 300 nm, the propagation constants for the two modes reach a common value of $k_\parallel = 14.8$ rad/μm, which coincides the value for a solitary single wire with the same cross section (red circled data points in Fig. 2). The gap size of 300 nm thus marks the distance at which the SPPs on two wires no longer couple.

For deeper investigations over the coupling between guided SPPs on individual wires, we have recorded the near-field intensity for a TWTL with a Y-split, as shown in Fig. 4. Either TE or TM mode is injected from the left side. The near-field intensity is recorded inside the gap for TE injection and 5 nm away from the outer surface of the wire for TM mode. With a +45 and -45 degree bend of each of the two wires, the TWTL splits into two parallel single nanowires separated with a distance of 610 nm, which ensures a negligible coupling between the wires. The two open ends of the nanowires then reflect the guided SPPs and result in standing wave patterns that allow for analyzing the effective wavelength and wavevector of the guided modes. It can be clearly seen that, regardless of the injected modes, after the Y-split the wavevector restores to the value of guided SPPs on single nanowires, i.e. $k_{\parallel,SPP} = 14.8$ rad/μm. Such simple structure may serve as a model system for the study of the coupling between guided SPPs. It is worth noting that our Y-split is not adiabatic and significant power loss is expected. To estimate the total loss caused by the Y-split, one needs to know the power reflection at the Y-split due to impedance mismatch and the power loss after the Y-split owing to the wire bending. Since different injected modes exhibit distinct characteristic impedance, they experience different degree of impedance mismatch at the Y-split. Analyzing the standing wave pattern on the TWTL for each mode should provide the information of power reflection [7]. Such analysis requires, however, the TWTL to be very long due to the long propagation length of the TM mode and is out of the scope of this work. Since the injected mode on the TWTL, *i.e.* TE or TM mode, is fully converted back to the guided mode on a single wire, the loss can be readily obtained by analyzing the bending loss of a single wire waveguide. In our design, the loss introduced by the 45 degree bending is optimized to about 0.2 dB.

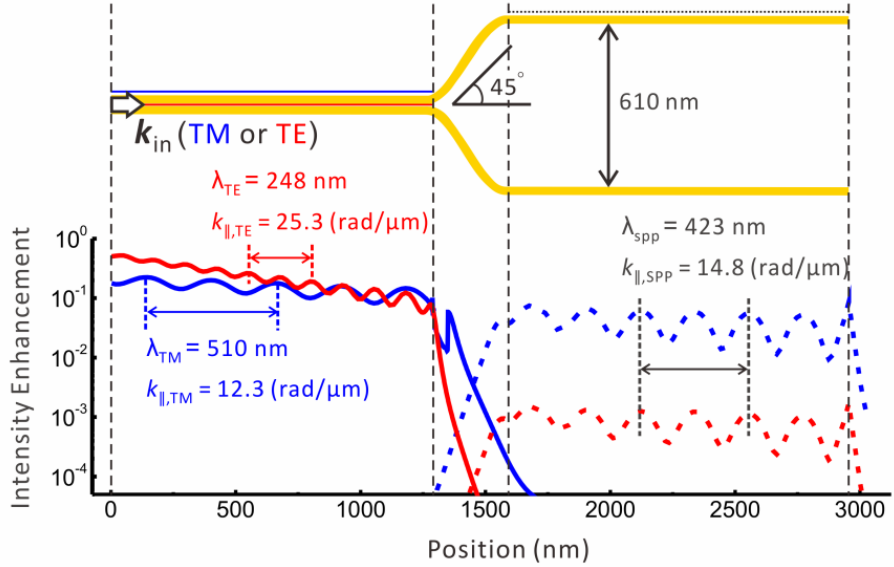

Fig. 4. Near-field intensity enhancement of the standing waves on a TWTL with a Y-split. The TWTL (cross section: 30×30 nm$^2$, gap: 10 nm) splits into two single wires separated by a distance of 610 nm. Either TE or TM mode is injected from the left side of the TWTL. Field intensity of the injected TM mode (blue solid line) and TE mode (red solid line) before the Y-split is recorded 5 nm away from the outer surface of the TWTL and at the center of the gap, respectively, as depicted with blue and red solid lines in the upper panel. After the Y-split, field intensity distributions is recorded 5 nm away from the surface of the wires, as depicted with dotted line in the upper panel. Propagation constant ($k_\parallel$) and the corresponding wavelength (λ) of each guided mode are marked. Regardless of the injected mode, after the Y-split, the wavevector and effective wavelength of the guided mode restore to the values of a single wire due to the vanishing near-field coupling.

## 3. Plasmonic mode converters

### 3.1. Conversion mechanisms

Having analyzed the coupling between the guided SPPs on solitary wires and the guided modes on a TWTL, now we may start constructing our mode converters. In order to switch between modes on a TWTL, additional phase difference between the current elements on individual wires needs to be introduced. The phase difference can be expressed as

$$\Delta\Phi = k_{\parallel,1} r_1 - k_{\parallel,2} r_2 , \qquad (4)$$

where the numbers at the subscript denote different wires of a TWTL. For two wires having identical material and cross sectional geometry embedded in the same medium, the propagation constants of the plasmonic modes on individual wires are identical and the additional phase difference is determined by the difference in the path length $\Delta r$, i.e.

$$\Delta \Phi = k_{\|} \Delta r . \qquad (5)$$

A conversion between TE and TM modes on a TWTL can thus be achieved by having one wire longer than the other by a length difference

$$\Delta r = \frac{m\pi - \theta_c}{k_{\|}} , \qquad (6)$$

where $m$ is an odd integer number and $\theta_c$ takes into account the phase offset introduced at the entrance and the exit of the converter. The top panel of Fig. 5(a) illustrates such kind of mode converter. Alternatively, the required phase difference can be introduced by modifying the propagation constant of the guided SPPs on individual wires. The required condition for successful mode conversion is

$$\Delta k_{\|} = \frac{m\pi - \theta_c}{R} , \qquad (7)$$

where $R$ is the effective length of the converter. Modification of the wavevector of the SPPs on individual wires can then be achieved by differing the wire cross sectional geometry (Fig. 5(a), middle panel) or the refractive index of the surrounding (Fig. 5(a), bottom panel). Note that we increased the gap size in the converter region in order to reduce the wire coupling such that propagation constant and length for a solitary wire (Fig. 2) can be directly used.

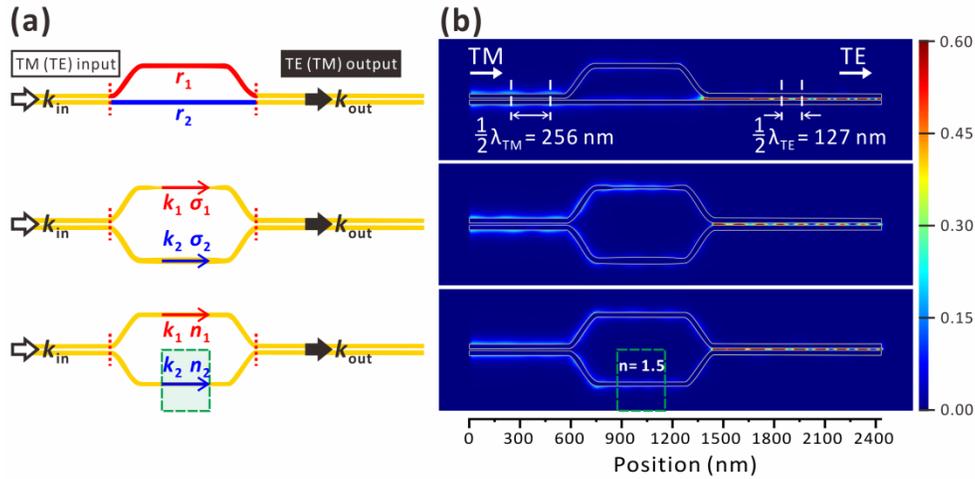

Fig. 5. (a) Schematic diagram of exemplary mode converters based on three working principles. Conversion of the input mode with propagation constant $k_{in}$ to the output mode with $k_{out}$ is accomplished through different path length ($r_1$ = 648 nm and $r_2$ = 500 nm, top panel) and different propagation constant ($k_1$ and $k_2$) due to different cross sectional width ($\sigma_1$ = 13 nm and $\sigma_2$ = 40 nm, middle panel) or different surrounding refractive index ($n_1$ = 1.0 and $n_2$ = 1.5, bottom panel). Green dashed square is the area in which the index of refraction is modified. The red dashed line indicates the position of the field monitor where modal profile and guided power flux are recorded. (b) Map of electric field intensity enhancement recorded at a frequency of 361.196 THz on a plane cutting the sample at the middle height. The phase difference between the displacement currents on individual nanowires is added an odd integer

multiples of π after the converter. The injected TM modes with most of the power guided through the outer surface of the wires are successfully converted into a highly concentrated TE mode in the dielectric nanogap (10 nm). All intensity maps are normalized to the same color scale.

Figure 5(b) plots the maps of the enhancement of electric field intensity on a plane cutting through the structure at the middle height. After passing through the mode converter, the loose field distribution of the injected TM mode is clearly switched to a concentrated TE mode in the nanogap. Here, we terminated the TWTL with an open end in order to build up a clear standing wave pattern for the analysis of the wavevector and impedance matching [7]. By analyzing the near-field intensity undulation, where the distance between two maxima corresponds to half of the wavelength of the guided mode, the propagation constant of guided mode can be obtained. As shown in the top panel of Fig. 5(b), the field inside the gap clearly reveals a distance of 127 nm between two maxima, corresponding to a propagation constant of 24.7 rad/μm, which is very close to the value for purely TE mode ($k_{\parallel,\text{TE}} = 25.3$ rad/μm). Although we have chosen to inject TM guided mode, all converters work well with TE injection. It is worth mentioning that the three tuning principles illustrated here work well independently but may also be combined to obtain higher flexibility. In view of practical use, mode converters using the difference in path length or cross sectional geometry require careful design and the conversion efficiency is fixed once the fabrication process is completed. In contrast, converter using the difference in the refractive index is more powerful since the index can be tuned actively by using active materials such as photoconductive semiconductors [43-45], phase transition materials [46, 47] and liquid crystals [48-50].

*3.2. Conversion efficiency*

In order to evaluate the conversion efficiency, we monitored the power flux on the outer wire surfaces and inside the nanogap to characterize the variation of the modal profile. We defined a mode characteristic

$$\text{MC} = \frac{P_{\text{total}} - 2P_{\text{gap}}}{P_{\text{total}}}, \tag{8}$$

where $P_{\text{total}}$ is the time-average total guided power and $P_{\text{gap}}$ is the power flux obtained by integrating the Poynting vector over an optimized cross sectional area $A_{\text{gap}}$ covering the dielectric gap and a small portion of metals. Note that, here, the TWTL extends out of the perfectly-matched layer boundaries such that the guided power is not reflected. A negative (positive) MC indicates that the guided power resides more inside (outside) the dielectric gap and the mode has more TE (TM) character. Ideally, MC would have a value of -1 for purely TE mode and +1 for purely TM mode. However, due to the partial overlap of the field distribution of TE and TM modes in space, the MC value never reaches the limits of ±1. The range of MC depends on the position and the size of $A_{\text{gap}}$. We have optimized $A_{\text{gap}}$ and obtained the largest range of MC (-0.46 and +0.82), meaning the best sensitivity to the variation of spatial modal distribution of the guided modes. Having obtained the MC, the modal conversion efficiency $\eta_{\text{m}}$ was then calculated using

$$\eta_{\text{m}} = \frac{\left|\text{MC}_{\text{out}} - \text{MC}_{\text{in}}\right|}{\left|\text{MC}_{\text{TE}}\right| + \left|\text{MC}_{\text{TM}}\right|}, \tag{9}$$

where $\text{MC}_{\text{TE}}$ and $\text{MC}_{\text{TM}}$ are the mode characteristics for purely TE and purely TM modes on the TWTL and $\text{MC}_{\text{in}}$ and $\text{MC}_{\text{out}}$ are values recorded before and after the mode converter on the TWTL. In addition to the modal profile, the power carried by the guided mode is also crucial for real applications. We, therefore, included a power transmission efficiency $\eta_p$ in the estimation of total conversion efficiency by calculating the ratio of total guided power before and after the converter using

$$\eta_p = \frac{P_{\text{total}}'}{P_{\text{total}}}, \quad (10)$$

where $P_{\text{total}}$ and $P_{\text{total}}'$ are the total power flux before and after the mode converters, respectively. The overall conversion efficiency of the mode converter can then be evaluated by

$$\eta_{\text{overall}} = \eta_p \cdot \eta_m. \quad (11)$$

The overall conversion efficiencies for the three exemplary converters shown in Fig. 5 are 35.4%, 22.4% and 21.5%, for the length-difference (top), cross section-difference (middle) and index-difference converter (bottom), respectively. The current metric for conversion efficiency is defined under the unique spatial mode distribution. Future direction in obtaining more accurate measures on the conversion efficiency may be performed in the Fourier-domain to associate the power related to each mode.

## 4. Applications in a complex integrated nanocircuit

Due to the distinct modal profile and propagation constant, the input impedance of the two modes on TWTL can be very different. One explicit example is the difference in the capability of driving optical nanoantennas. TE modes are capable of driving the bright mode of a linear gap nanoantenna from the central feed gap and the near-field optical energy can then be re-emitted to the free space as propagating photons through the antenna [7, 41]. In contrast, the guided TM mode will be strongly reflected since its impedance matches only to the dark antenna mode, with which the localized energy cannot be converted into propagating photons [41]. From this viewpoint, our converters offer possibility to control the radiation from a nanoantenna by manipulating the optical impedance of the guided power.

As an example for the application of our mode converter, we show controlled and improved power transmission in a complex integrated nanocircuit consisting of a single metal stripe in connection with a TWTL followed by an index mode converter and terminated by an optimized nanoantenna as a near-to-far-field coupler, as depicted in Fig. 6(a). Single metal stripes and nanowires are commonly used waveguides for SPPs [18, 20, 21, 51]. However, the mismatch in the modal distribution makes it hard to excite the radiative mode of a gap nanoantenna using single metal stripe. In the following, we show that using our mode converter, "unbalanced" optical energy guided on a single metal stripe can be used to excite radiative antenna mode and the optical power transmitted from the near to the far field can be controlled. Our mode converter, therefore, serves as a plasmonic counter part of "balun", which improves the impedance matching between "unbalanced" signals guided by the single metal stripe and "balanced" signals on the TWTL. We demonstrate mode conversion by an index mode converter because this converter can be actively controlled and is of potential for index sensing. The cross sectional geometry of the single metal stripe is the same as that of the TWTL, except that the gap is filled up by gold. The total length of the nanoantenna has been optimized to achieve best radiation efficiency.

Let us start with the transmission of the guided power from the single metal stripe to the TWTL. Considering the modal distribution of the guided SPPs on single metal stripe, it is obvious that the TM mode on the TWTL matches guided SPPs much better than the TE mode does. The junction between single metal stripe and TWTL indeed serves as a mode filter, which only allows transmission of TM mode on TWTL and SPPs on a single metal stripe. As a result, the optical power from the single metal stripe couples only to the TM mode on the TWTL, whereas the TE mode cannot be excited due to impedance mismatch. After successful transmission of power to the TWTL, the guided TM mode then passes through the mode converter. With same path length and cross section of each wire in the converter, the conversion efficiency is completely determined by the refractive index of the surrounding media. If the refractive index of the material inside the green box area is identical to that of the rest surrounding, *i.e.* $n = 1.0$, there would be no conversion of mode and the guided power will be reflected at the nanoantenna terminal since the TM mode cannot drive the radiative antenna mode. Figure 6(b) clearly shows that the guided TM optical power is reflected back to the nanocircuit and a pronounced standing wave pattern is generated around the outer surface of the TWTL. With refractive index inside the green box swept from 1.0 to 2.0, the guided TM mode is gradually converted into TE mode and the emitted power therefore increases and reaches a maximum at $n = 1.5$, corresponding to the highest overall conversion efficiency. Figure 6(c) shows the intensity map of the circuit with the refractive index in the green box set to 1.5. From the modal distribution recorded at different sections of the circuit, shown in Fig. 6(d), it is obvious that the TM mode is successfully converted into TE mode when $n = 1.5$ and the radiative mode of the nanoantenna is successfully excited. Following the impedance matching argument and according to the reciprocity theorem [6], it becomes obvious that the bonding and anti-bonding mode of a symmetric gap nanoantenna [41] can be used to efficiently excite the guided TE and TM modes on a plasmonic TWTL, respectively.

Since the radiative mode of the nanoantenna can only be driven by the TE mode, the emitted power by the antenna can be expressed as

$$P_{out} = \eta_{antenna} \times \left(1-\Gamma_{p2}\right) \times e^{-2k'_{\parallel,TE}r_3} \times \eta_{overall} \times e^{-2k'_{\parallel,TM}r_2} \times \left(1-\Gamma_{p1}\right) \times P_S, \qquad (12)$$

where $P_S$ is the injected power, $\Gamma_{p1}$ is the power reflectivity at the junction between single stripe and TWTL, $\Gamma_{p2}$ is the power reflectivity at the TWTL-to-antenna junction and $\eta_{antenna}$ is the radiation efficiency of the nanoantenna [7]. It is clear that the emission power of the nanoantenna is a linear function of the overall conversion efficiency. Since $\eta_{overall} = \eta_p \cdot \eta_m$, the emission power is controlled by $\eta_m$. Figure 6(e) plots the mode conversion efficiency $\eta_m$, power conversion efficiency $\eta_p$ and the overall conversion efficiency $\eta_{overall}$, along with the emission power $P_{out}$ from the optimized nanoantenna as a function of the refractive index of the material inside the green box. It can be seen that the power conversion efficiency is relatively constant with respect to the change of surrounding refractive index and the radiated power follows the trend of $\eta_m$. Taking the slope $s$ of the linear relationship between the wavevector and the index of refraction, as shown in Fig. 2(b), the mode conversion efficiency can be expressed as a function of the difference in refractive index of media surrounding the two wires in the converter area

$$\eta_m = \frac{1}{2}\left[\cos\left(\Delta k_\parallel L_c + \pi\right) + 1\right] = \frac{1}{2}\left[\cos\left(s\Delta n L_c + \pi\right) + 1\right]. \qquad (13)$$

Therefore, the emission power of the antenna can be expressed as a cosine function of the index difference and the sensitivity for index sensing is proportional to $sL_c$. In principle, the sensitivity can be improved by using smaller wire cross section or longer converter. However, the effect of increasing loss for longer waveguide needs to be taken into account in order to obtain optimized geometry. It is worth mentioning that the control over the antenna radiation can be made active by, for example, covering one of the wires with a photoconductive material or with well-designed microfluidic channels such that deterministic difference in the refractive index can be actively introduced.

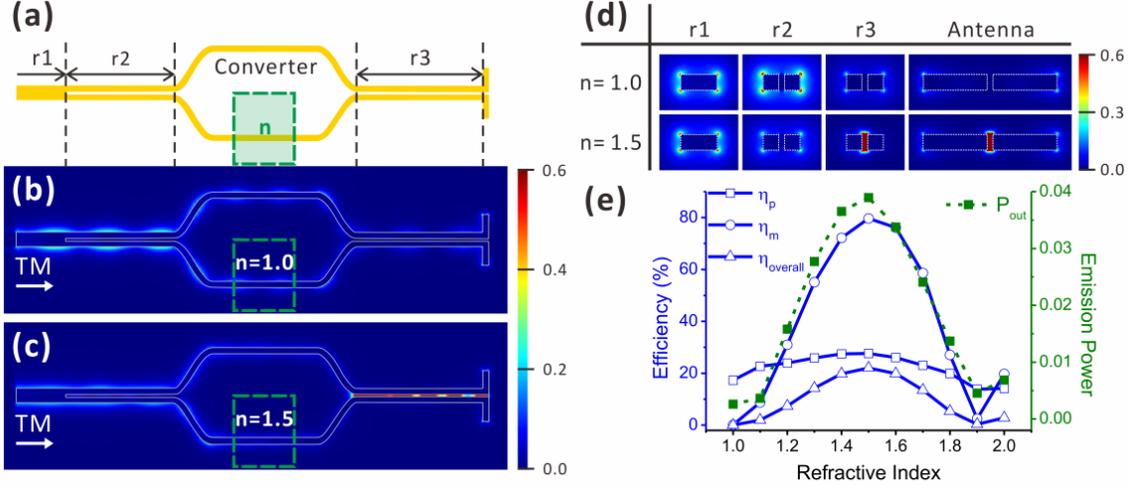

Fig. 6. (a) Schematic diagram of a complex integrated plasmonic nanocircuit consisting of a single metal stripe (r1), the first section of TWTL (r2), a mode converter, the second section of TWTL (r3) and a nanoantenna. The cross section of the wires of TWTL and nanoantenna arms is 30×30 nm$^2$ and the gap is 10 nm. The cross section of the single metal stripe is 30×70 nm$^2$, exactly the same as that of the TWTL except that the gap is filled with gold. The surrounding index of refraction is 1.0 everywhere except for the tuning area indicated by the green dashed box, in which the refractive index is scanned from 1.0 to 2.0. (b) and (c) the field intensity enhancement map of the circuit obtained with a refractive index set to 1.0 and 1.5, respectively, corresponding to the case of no conversion and best conversion. (d) Cross sectional profile of field intensity enhancement at different section of the complex circuit and at the terminal antenna obtained at $n = 1.0$ and $n = 1.5$. (e) Mode ($\eta_m$, open circles), power ($\eta_p$, open squares) and overall conversion efficiency ($\eta_{overall}$, open triangles) as well as the emitted power from the nanoantenna ($P_{out}$, solid squares dashed line) as a function of the refractive index of the material inside the green dashed box. The radiated power is normalized to the injected power. All simulation data are obtained at a frequency of 361.196 THz.

## 5. Perspective on nanoscale light-matter interaction

Plasmonic nanostructures show general similarities to natural chemical molecules [52] and the optical impedance of nanoantennas and quantum emitters can be described under an unified framework [53]. As we have demonstrated, using TWTL and our mode converter, we may control the emission of a nanoantenna by controlling the optical impedance of the guided power. Replacing the nanoantenna with a quantum emitter that has its own optical impedance, such as a single molecule, we may, in principle, manipulate the light-matter interaction in a similar way. In view of nanoscale light-matter interaction, the much reduced mode volume $V_{mode}$ substantially enhances the coupling strength $g \propto [V_{mode}]^{-1/2}$ between the nanocavity and single quantum systems [3]. In addition, the well-defined polarization of the electric field may further improve the interaction between the guided field and the quantum systems residing in the nanogap. From classical point of view, the energy of a molecule with dipole moment $\boldsymbol{\mu}$ in a polarized electric field $\boldsymbol{E}$ is just their dot product,

$$\boldsymbol{E} \cdot \boldsymbol{\mu} = |\boldsymbol{E}||\boldsymbol{\mu}|\cos\phi, \qquad (14)$$

where $\phi$ denotes the angle between the two vectors. If the electric field and the dipole moments can be aligned such that $\phi = 0$, the light-matter interaction can be further enhanced. From experimental point of view, the alignment of quantum systems may be accomplished through the use of thin crystalline host films [54, 55]. Using recently proposed hybrid fabrication approach for atomically smooth plasmonic nanostructure [56] and the "stacked" geometry for extremely small gap [57], it is possible to integrate thin layer of various materials, including gain, liquid crystal, photoconductive semiconductor or graphene to enable multiple functions of nanocircuits.

## 6. Conclusion

We have presented realizable plasmonic mode converters to manipulate the guided modes on a plasmonic TWTL. The mode conversion is accomplished through the introduction of additional phase difference between guided SPPs on individual wires by differing the path length, the cross section, and by changing the surrounding index of refraction. We analyzed the conversion efficiency of the mode converter and demonstrated its capability in manipulating the optical impedance of guided mode and therefore the emission power of a nanoantenna in a complex nanocircuit. Being able to manipulate the guided mode is important for the control over guiding properties and optical impedance of the mode and provides possibility to achieve active manipulation of the interaction between optical field and nanoobjects. One may think about building up a nanocavity consisting of a piece of TWTL terminated by two gap nanoantennas, the capability to control the mode and near-to-far field conversion efficiency then means the ability to tune the reflectivity and thus the quality factor of the cavity. Our converter may find interesting applications in optical impedance control, nanosensors and nanoresonators. We anticipate a number of applications in optical nanocircuits and active control of the light-matter interaction at the nanoscale.

## Acknowledgements

The work was supported by the National Science Council of Taiwan under grants NSC 99-2113-M-007-020-MY2 and NSC 100-2112-M-007-007-MY3. J.-S. H. thanks C.-S. Huang and Dr. T. Lankau for valuable discussions.